\RequirePackage{lineno}
\documentclass[12pt,a4paper]{article}

\usepackage{cancel}
\usepackage{cite}
\usepackage{graphicx}
\usepackage{epsfig}
\usepackage{amsmath,amsfonts,amssymb}
\usepackage{t1enc}

\newcommand{\oh}{\frac{1}{2}}
\newcommand{\afb}{A_{FB}}
\newcommand{\afbl}{A_{FB}^\ell}
\newcommand{\afbll}{A_{FB}^{\ell \ell}}

\newcommand{\ttb}{t \bar t}

\newcommand{\ppb}{p \bar p}

\begin{document}
\begin{flushright}
CERN-PH-TH-2014-081
\end{flushright}

\begin{center}
\begin{Large}
{\bf Quantum coherence, top transverse polarisation \\[1mm] and the Tevatron asymmetry $A_{FB}^\ell$}
\end{Large}
\vspace{2mm}

J. A. Aguilar--Saavedra \\[1mm]
\begin{small}
{\it Departamento de F\'{\i}sica Te\'orica y del Cosmos, Universidad de Granada, \\ E-18071 Granada, Spain.} \\
{\it PH-TH Department, CERN, CH-1211 Geneva 23, Switzerland.}
\end{small}
\end{center}

\begin{abstract}
We revisit the relation between the asymmetries $A_{FB}$ and $A_{FB}^\ell$ in $t \bar t$ production at the Tevatron, using as new physics benchmark a colour octet. We find that $A_{FB}^\ell$ receives large contributions from the interference between $\lambda = \pm 1/2$ top helicity states, which has been ignored in some of the previous literature on the subject. The omission of these contributions results in a severe underestimation of the asymmetry, around $1/2$ and $1/50$ of the true value for right-handed and left-handed top couplings to the octet, respectively. Interference effects are closely related to a sizeable transverse top polarisation, as yet not considered in this context.
\end{abstract}

Since some time, the CDF and D0 experiments have found anomalies in the measurement of several forward-backward (FB) asymmetries in $\ttb$ production at the Tevatron (see~\cite{Aguilar-Saavedra:2014kpa} for a recent review). The largest deviations with respect to the Standard Model (SM) predictions were found in the $\ttb$ production asymmetry~\cite{Aaltonen:2012it,Abazov:2014cca},
\begin{equation}
\afb = \frac{N(\Delta y >0) - N(\Delta y <0)}{N(\Delta y >0) + N(\Delta y <0)} \,,
\label{ec:afb1}
\end{equation}
with $\Delta y = y_t-y_{\bar t}$ the difference between the top and antitop rapidities in the laboratory frame. (The asymmetry is the same when the rapidities are taken in the $\ttb$ rest frame.) A second asymmetry involves the rapidities of the charged leptons $\ell$ produced in the semileptonic decay of top (anti-)quarks $t \to W b \to \ell \nu b$~\cite{Aaltonen:2013vaf,Aaltonen:2014eva,Abazov:2014oea,Abazov:2013wxa},
\begin{equation}
\afbl = \frac{N(q_\ell y_\ell >0) - N(q_\ell y_\ell <0)}{N(q_\ell y_\ell >0) + N(q_\ell y_\ell <0)} \,,
\end{equation}
with $y_\ell$ the rapidity of the lepton and $q_\ell$ its charge. A third asymmetry $\afbll$ is also measured when both quarks decay semileptonically, but its statistical uncertainty is larger, and will not be considered here. For $\afb$, the CDF Collaboration reports $\afb = 0.164 \pm 0.045$, which is $1.7\sigma$ above the SM prediction $\afb^\text{SM} = 0.088$~\cite{Bernreuther:2012sx}, and the D0 Collaboration measures $\afb = 0.106 \pm 0.030$, compatible with the SM. The naive average of these two values and $\afb = 0.42 \pm 0.16$ in the dilepton channel~\cite{AFBCDFdil} gives $\afb = 0.131 \pm 0.024$, which is $1.7\sigma$ above the SM value. For the lepton asymmetry, the average of CDF and D0 results gives $\afbl = 0.069 \pm 0.019$, $1.6\sigma$ above the SM prediction $A_{FB}^{\ell,\text{SM}} = 0.038$~\cite{Bernreuther:2012sx}.

The observation of deviations in the two asymmetries, which were larger in previous measurements~\cite{Aaltonen:2011kc,Abazov:2011rq}, has fuelled the study of their interrelation, in order to test different new physics explanations for the anomalies~\cite{Falkowski:2011zr,Berger:2012nw,Berger:2012tj,Falkowski:2012cu,Carmona:2014gra}, as well as to check the SM prediction for their ratio~\cite{Falkowski:2012cu,Carmona:2014gra}.
Motivated by some discrepancy between results of~\cite{Berger:2012nw,Berger:2012tj} and~\cite{Falkowski:2011zr,Falkowski:2012cu,Carmona:2014gra},\footnote{Most conspicuously, the hierarchy $\Delta \afbl \lesssim \Delta \afb$, derived in~\cite{Berger:2012nw} for the new physics contributions to the asymmetries, is violated in several benchmark points of~\cite{Falkowski:2011zr,Falkowski:2012cu,Carmona:2014gra}.} in this Letter we revisit the relation between $\afbl$ and $\afb$ and investigate the effect of quantum interference between top helicity states, not taken into account in the derivations of~\cite{Berger:2012nw,Berger:2012tj}.

For our study, we consider a benchmark model of a {\it light} colour octet~\cite{Barcelo:2011vk,Tavares:2011zg,Alvarez:2011hi,AguilarSaavedra:2011ci,Krnjaic:2011ub} with a large width in order to comply with the constraints from dijet pair production~\cite{Gross:2012bz,Gresham:2012kv}. 
Apart from being the model that gives best agreement with all $\ttb$ data~\cite{Aguilar-Saavedra:2013rza}, a colour octet allows to explore the relation between $\afbl$ and $\afb$ in various scenarios, since the chirality of the octet coupling to the light quarks $q=u,d$ and to the top quark is almost arbitrary. (We do not consider constraints from $B$ physics, which are not important for the size of the couplings considered here~\cite{Bai:2011ed,Haisch:2011up}.) The relevant interaction Lagrangian is~\cite{AguilarSaavedra:2011vw}
\begin{equation}
\mathcal{L} = - \left[ \bar u \gamma^\mu {\textstyle \frac{\lambda^a}{2}} (g_V^u + \gamma_5 g_A^u) u 
+  \bar d \gamma^\mu {\textstyle \frac{\lambda^a}{2}} (g_V^d + \gamma_5 g_A^d) d 
+  \bar t \gamma^\mu {\textstyle \frac{\lambda^a}{2}} (g_V^t + \gamma_5 g_A^t) t 
\right] G_\mu^a  \,.
\end{equation}

The model is implemented in the generator {\sc Protos}~\cite{AguilarSaavedra:2008gt} that calculates the tree-level matrix element for the $2 \to 6$ processes involved in $\ttb$ production and subsequent decay $t \bar t \to W^+ b W^- \bar b \to f_1 \bar f_1' b \bar f_2  f_2' \bar b$, with $f_i \bar f_i' = u\bar d, c\bar s, \ell \bar \nu$, keeping all spin information in the decay chain. As a cross-check, we select three benchmark points of~\cite{Falkowski:2012cu}, with an octet mass $M=200$ GeV and width $\Gamma = 50$ GeV, finding the new physics contributions to the asymmetries
\begin{align}
& g_R^{u,d,t} = 0 \,,\quad g_L^{u,d,t} = 0.8 g_s \,: && \Delta \afbl = -0.07 \,, \notag \\
& g_R^{u,d,t} = 0.8 g_s \,,\quad g_L^{u,d,t} = 0 \,: && \Delta \afbl = 0.16 \,, \notag \\
& g_R^{u,d,t} = 0.4 g_s \,,\quad g_L^{u,d,t} = -0.4 g_s \,: && \Delta \afbl = 0.05 \,,
\end{align}
with $\Delta \afb = 0.12$ in all cases, in good agreement with~\cite{Falkowski:2012cu}. We note that the total asymmetries are obtained, to a good approximation, by adding to these values the SM contributions $\afb^\text{SM} = 0.088$, $A_{FB}^{\ell,\text{SM}} = 0.038$. We will not include them since they do not affect our discussion and only amount to a shift of the results presented, and will instead concentrate on the new physics contributions $\Delta \afb$, $\Delta \afbl$.

\begin{figure}[htb]
\begin{center}
\includegraphics[height=7cm,clip=]{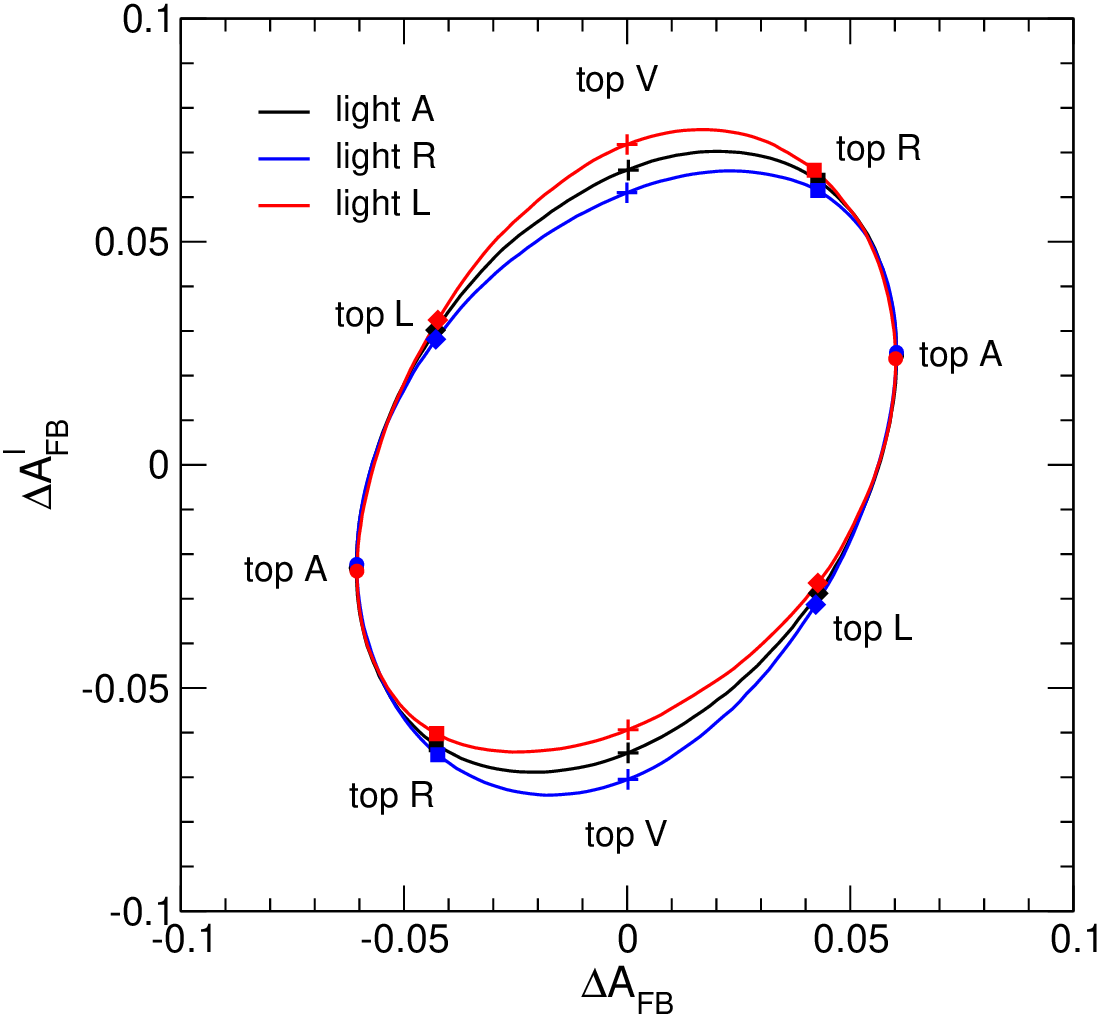}
\end{center}
\caption{Relation between $\Delta \afb$ and $\Delta \afbl$, for the three choices of light quark couplings (A/R/L) and top couplings given by Eq.~(\ref{ec:gtop}). The points corresponding to axial (A), vector (V), left (L) and right (R) couplings of the top quark are indicated. }
\label{fig:A-good}
\end{figure}

Our exploration of the octet parameter space is done for $M=250$ GeV and $\Gamma/M = 0.2$. It is assumed for simplicity that up and down quarks have the same couplings, $g_V^u = g_V^d$, $g_A^u = g_A^d$. We fix $g_A^u [ (g_V^t)^2 + (g_A^t)^2 ]^{\oh} = 0.1$, with $g_A^u > 0$, being the overall sign of the octet contribution determined by the top coupling. We restrict ourselves to couplings to $u$, $d$ that are either axial, right-handed or left-handed, and scan over all possible chiralities for the top quark couplings, parameterised as~\cite{Aguilar-Saavedra:2014nja}
\begin{align}
& \frac{g_A^t}{[ (g_V^t)^2 + (g_A^t)^2 ]^{\oh}} \equiv \cos \phi_h \,, && \frac{g_V^t}{[ (g_V^t)^2 + (g_A^t)^2 ]^{\oh}} \equiv \sin \phi_h \,,
\label{ec:gtop}
\end{align}
with $\phi_h \in [0,2\pi]$. The asymmetries $(\Delta \afb,\Delta \afbl)$ so obtained are presented in Fig.~\ref{fig:A-good}.
The sign of $\Delta \afbl$ can be understood from the threshold behaviour. We reproduce here the argument in~\cite{Falkowski:2011zr}. Initial $q_R \bar q_R$ pairs have their spins aligned in the proton direction $\hat p$, therefore they have a total spin $\vec S \cdot \hat p = 1$. Their orbital angular momentum in this direction is zero, so the total angular momentum is $\vec J \cdot \hat p = 1$. At the threshold, the $\ttb$ pair is produced with zero orbital angular momentum, so angular momentum conservation implies $\vec S \cdot \hat p = 1$, that is, both spins in the proton direction independently of the production angle. Since the positive charge lepton from the top decay tends to follow the top spin direction (see the Appendix), it is preferentially emitted with $y_{\ell^+} > 0$. The negative charge lepton from the top decay tends to be emitted opposite to the top spin, so $y_{\ell^-} < 0$ and $q_{ell^-} y_{\ell^-} > 0$. For initial $q_L \bar q_L$ states the argument is the opposite, so that
$q_\ell y_{\ell} < 0$. For equal $q_R \bar q_R$ and $q_L \bar q_L$ cross sections, $\afbl = 0$. Note that this argument, valid strictly only at the threshold, does not depend on $\afb$.

Now let us turn to Fig.~\ref{fig:A-good}. The point labelled `top R' with $\Delta \afb > 0$ has $g_R^t > 0$, $g_L^t = 0$, so as to have $g_A^t > 0$ since we taking $g_A^u > 0$. The interference between the SM and octet amplitudes is proportional to $g_V^u g_V^t$ times a positive factor. For right-handed couplings to $u,d$, $g_R^u > 0$, $g_L^u = 0$, the positive SM--octet interference generates an excess of $q_R \bar q_R$ and $\Delta \afbl > 0$. For left-handed couplings to $u,d$, $g_R^u = 0$, $g_L^u < 0$ (since $g_A^u > 0$) the SM--octet interference decreases the $q_L \bar q_L$ cross section, and again $\Delta \afbl > 0$. For axial couplings $g_R^u > 0$, $g_L^u < 0$, there is an increase of $q_R \bar q_R$ and a decrease of $q_L \bar q_L$. For the point labelled `top L' one has $g_R^t = 0$, $g_L^t < 0$ and the argument is reversed: the SM--octet interference yields a depletion of $t_R \bar t_R$, an increase of $t_L \bar t_L$, or both.

Once the sign of $\Delta \afbl$ is well understood, there are several interesting conclusions to be drawn from Fig.~\ref{fig:A-good}.
\begin{enumerate}
\item For a definite sign of $\Delta \afb$, the relation between $\Delta \afbl$ and $\Delta \afb$ mainly depends on the chirality of the top quark coupling, parameterised by $\cos \phi_h$. Right-handed couplings lead to larger $\Delta \afbl$ than left-handed ones, for which $\Delta \afbl$ and $\Delta \afb$ can even have opposite signs. This fact is explained by the above discussion, and does not have any relation with the top helicity. For example, an octet with right-handed couplings to the top can induce a negative top polarisation $P$ in the helicity basis, as we will see in the following. The apparently reasonable argument that $P > 0$ leads to larger $\Delta \afbl$ and $P < 0$ to smaller $\Delta \afbl$ is simply not true. $\afbl$ also depends on the coupling to the light quarks to a smaller extent.
\item The hierarchy $\Delta \afbl \lesssim \Delta \afb$ does not hold, not even when both asymmetries have the same sign. Noticeably, a large $\Delta \afbl$ is possible even for zero $\Delta \afb$, when the top coupling to the octet is vectorial.
\item A $\ttb$ asymmetry above the SM value, $\Delta \afb > 0$, is compatible with positive, negative, or vanishing $\Delta \afbl$.
\item The current averaged values of both asymmetries are fitted by $\Delta \afb = 0.043$, $\Delta \afbl = 0.031$, which in the octet model would correspond to a top coupling between axial and right-handed~\cite{Aguilar-Saavedra:2014nja}.
\end{enumerate}

Now we turn our attention to the effect of helicity interference, not taken into account in~\cite{Berger:2012nw,Berger:2012tj}.
Top quarks are in general produced in a (spin) state that can be described by a $2 \times 2$ Hermitian density matrix. Setting a coordinate system $(x,y,z)$ in the top quark rest frame, the density matrix reads
\begin{equation}
\rho=\frac{1}{2}\left( \! \begin{array}{cc} 1+P_z & P_x-i P_y \\ P_x+i P_y & 1-P_z
\end{array} \! \right) \,,
\label{ec:rho}
\end{equation}
where $P_i = 2 \langle S_i \rangle$, with $i=x,y,z$, using the basis $\{|+\rangle,|-\rangle\}$ where $S_z$ is diagonal. The three polarisations are denoted as `longitudinal' ($P_z$), `transverse' ($P_x$) and `normal' ($P_y$). There are two situations in which one can sum over longitudinal polarisations incoherently, that is, assume that a fraction $(1+P_z)/2$ of top quarks is produced in a pure state $|+\rangle$ with spin component $+1/2$ in the $\hat z$ direction and a fraction $(1-P_z)/2$ is produced in a state $|-\rangle$ with spin component $-1/2$. Incoherent sums can be performed, obviously, if the off-diagonal entries in $\rho$ vanish because of our choice of the $\hat z$ axis --- note that a Hermitian matrix can always be diagonalised. But off-diagonal elements do not necessarily vanish when using the helicity basis, that is, choosing the $\hat z$ axis as the top momentum in the $\ttb$ CM frame $\vec p_t$. The second situation that allows for incoherent sums is when the observables considered are independent of the azimuthal angle $\phi$ of the $W$ boson momentum in the $(x,y,z)$ reference system, which can then be trivially integrated~\cite{AguilarSaavedra:2012xe}. This is the case for $\afb$, but obviously not for $\afbl$.

\begin{figure}[htb]
\begin{center}
\includegraphics[height=7cm,clip=]{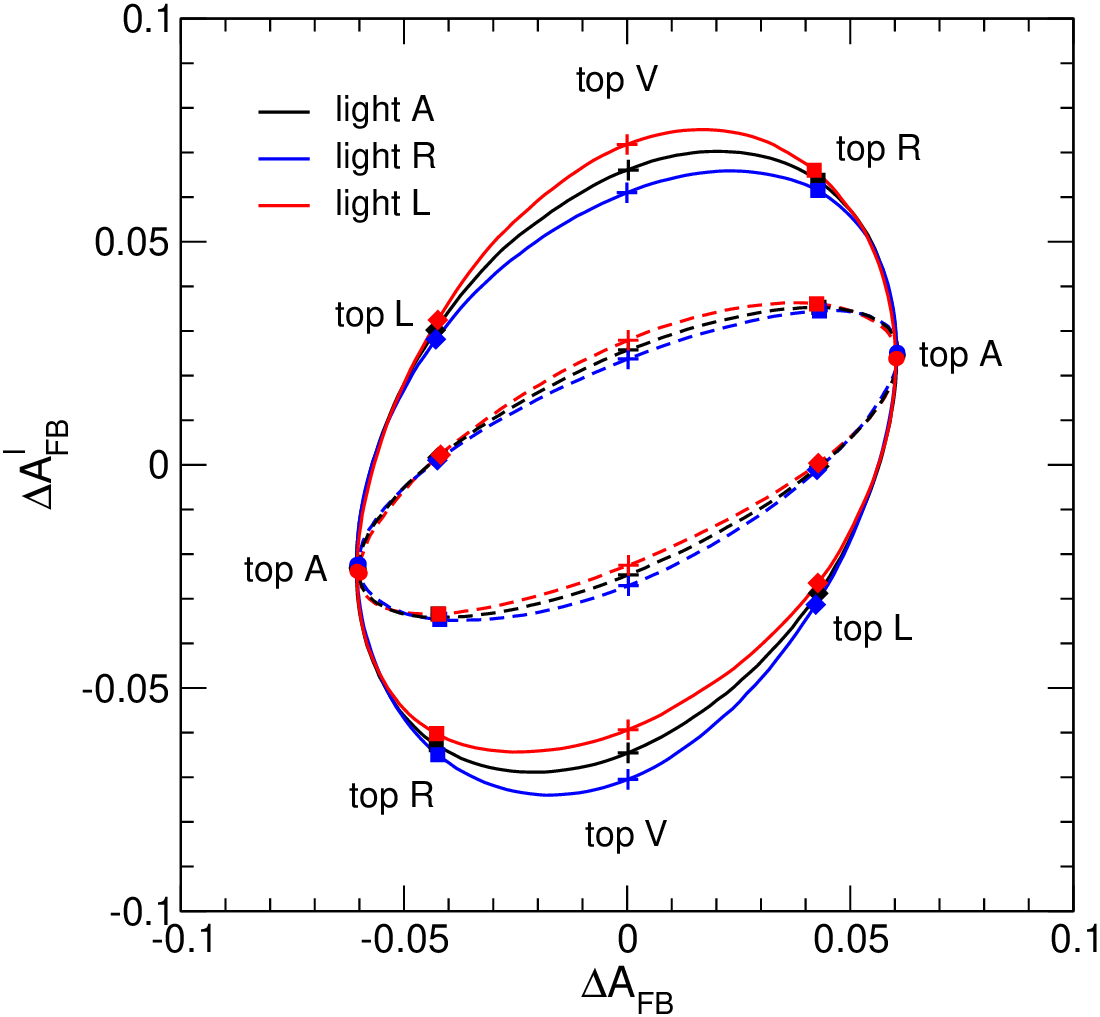}
\end{center}
\caption{Relation between $\Delta \afb$ and $\Delta \afbl$ with top helicity interference (solid lines) and without it (dashed lines).}
\label{fig:A-bad}
\end{figure}

The influence of helicity interference in the generated $\afbl$ is investigated by implementing the helicity projectors in {\sc Protos}. The charged lepton distributions in the top quark rest frame confirm that the projectors indeed yield pure helicity states (see the Appendix). The no-interference asymmetries $\Delta A_{FB}^0$, $\Delta A_{FB}^{\ell,0}$ are obtained as~\cite{Berger:2012nw,Berger:2012tj}\footnote{An overall $\mathcal{O}(1)$ cross section normalisation factor applied in those references, common to both asymmetries, is dropped here to keep consistence with the results presented above, and since it does not affect our discussion.}
\begin{align}
& \Delta A_{FB}^{0} = \frac{\sigma_+ \Delta A_{FB}^+ + \sigma_- \Delta A_{FB}^-}{\sigma_+ + \sigma_-} \,, && 
\Delta A_{FB}^{\ell,0} = \frac{\sigma_+ \Delta A_{FB}^{\ell,+} + \sigma_- \Delta A_{FB}^{\ell,-}}{\sigma_+ + \sigma_-} \,,
\label{ec:aaprox}
\end{align}
where the quantities indicated with plus (minus) signs are computed for top quarks of helicity $\lambda = 1/2$ ($\lambda = -1/2$). The results are presented in Fig.~\ref{fig:A-bad}.
As expected, $\Delta A_{FB}^0$ coincides with $\Delta A_{FB}$ since the $\ttb$ asymmetry, as well as the total cross section, is independent of $\phi$. On the other hand, the approximation in Eqs.~(\ref{ec:aaprox}) drastically underestimates $\Delta \afbl$, except if the top coupling is axial. For example, for right-handed top couplings, $\Delta A_{FB}^{\ell,0} / \Delta \afbl \simeq 0.55$, and for a left-handed ones $\Delta A_{FB}^{\ell,0} / \Delta \afbl = 0.02$. (As it is well known, for the massive top quark the chirality and helicity states do not coincide.) Most likely, a sizeable $\lambda=\pm 1/2$ interference is not a particular feature of the colour octet considered here, but it is also expected for other models proposed to explain the $\afb$ measurements where the top coupling is chiral, as for example new $Z'$ or $W'$ bosons. We also note that the importance of the interference is enhanced by the fact that often $\Delta A_{FB}^{\ell,+}$ and $\Delta A_{FB}^{\ell,-}$ have opposite signs and their contributions cancel.

\begin{figure}[htb]
\begin{center}
\includegraphics[height=7cm,clip=]{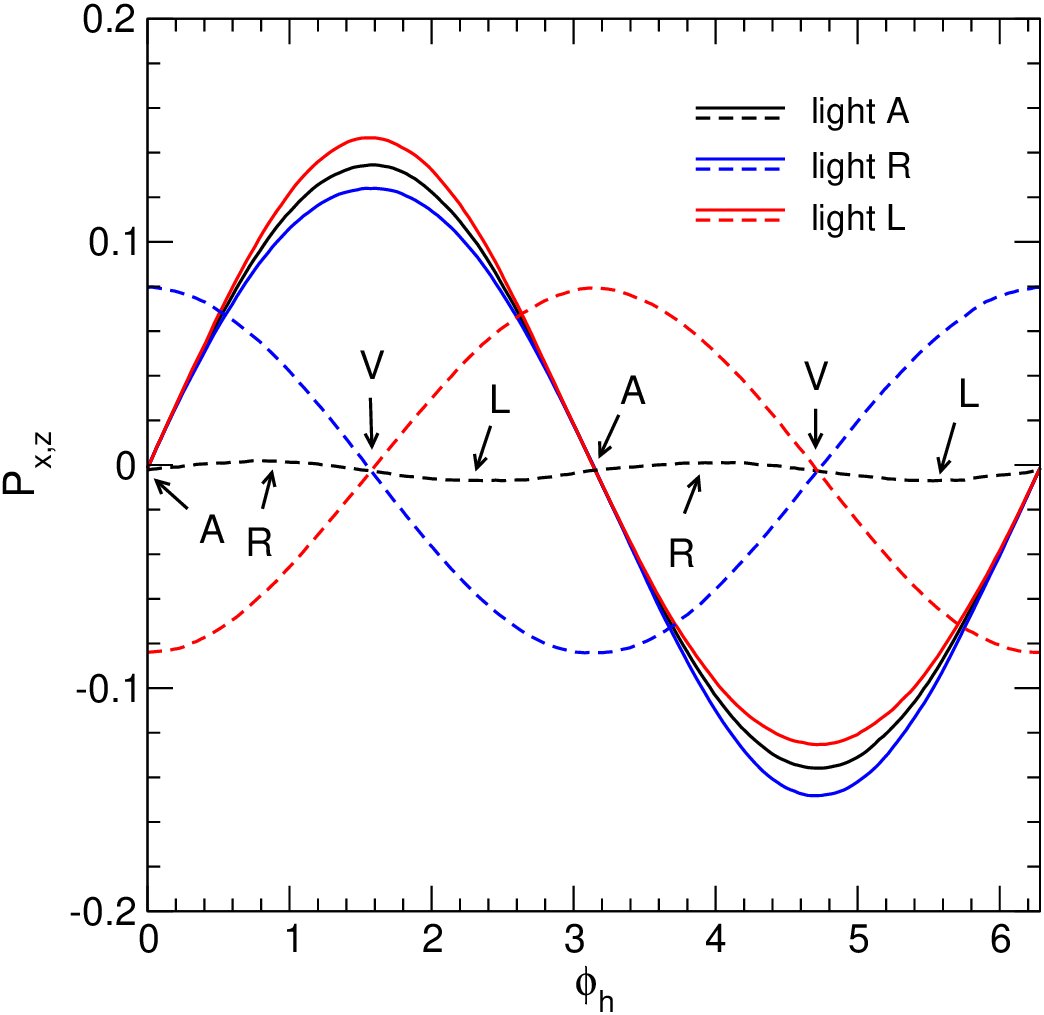}
\end{center}
\caption{Transverse (solid lines) and longitudinal (dashed lines) top polarisation as a function of the angle $\phi_h$ in Eq.~(\ref{ec:gtop}) that parameterises the chirality of the top coupling. Points corresponding to top axial (A), vector (V), right-handed (R) and left-handed (L) couplings are indicated.}
\label{fig:pol}
\end{figure}

The presence of helicity interference --- that is, the non-diagonal terms in the density matrix (\ref{ec:rho}) --- is associated to a large polarisation in a direction perpendicular to the helicity axis $\hat z$. We specify the other two directions by choosing $\hat y$ orthogonal to the production plane, and determine $\hat x$ by requiring that the coordinate system is right-handed. Specifically,
\begin{equation}
\hat z = \frac{\vec p_t}{|\vec p_t|} \,,\quad
\hat y = \frac{\vec p_t \times \vec p_p}{|\vec p_t \times \vec p_p|} \,,\quad
\hat x = \hat y \times \hat z \,,
\label{ec:axes}
\end{equation}
with $\vec p_p$ the proton momentum in the top quark rest frame. The polarisations in the three directions $\hat x$, $\hat y$, $\hat z$ can be determined by suitable angular asymmetries~\cite{Aguilar-Saavedra:2014eqa} involving the angle between the charged lepton and the corresponding axis. The so-called `transverse' polarisation in~\cite{Baumgart:2013yra} corresponds to the normal polarisation $P_y$ in this work, and is small in our case because the colour octet is lighter than the $\ttb$ threshold and the complex phase given by the octet propagator, required to generate a non-zero $P_y$, is small.
The results for $P_x$ and $P_z$ are presented in Fig.~\ref{fig:pol}, for the three chiralities for $u,d$ couplings considered, and as a function of the angle $\phi_h$. They deserve a detailed discussion.

For left- and right-handed couplings to $u,d$, a longitudinal polarisation $P_z$ arises from the interference between the SM and octet amplitudes, which is proportional to $g_V^q g_V^t$ as mentioned before. For example, for $g_R^u > 0$, $g_L^u = 0$ (remember that we take equal couplings to $u$ and $d$ and fix $g_A^u > 0$) a positive polarisation $P_z > 0$ can arise for $g_R^t > 0$, $g_L^t = 0$, so that the SM--octet interference produces an excess of $t_R$. But it can as well result from $g_R^t = 0$, $g_L^t < 0$, when the SM--octet interference produces a depletion of $t_L$. And in both cases $\Delta \afb > 0$, since $g_A^t > 0$. For left-handed couplings to $u,d$, $g_L^u < 0$, $g_R^u = 0$, the behaviour is the opposite. A top coupling $g_R^t > 0$, $g_L^t = 0$ produces $P_z < 0$, the same as $g_R^t = 0$, $g_L^t < 0$, and in the two cases $\Delta \afb > 0$.

For axial coupling to $u,d$ the SM--octet interference is zero and $P_z$ arises solely from the octet quadratic term, following the expectation: $P_z > 0$ for a right-handed top coupling, $P_z < 0$ for a left-handed one, and $P_z = 0$ for vector or axial couplings. Since the quadratic term is suppressed by the small couplings, the generated polarisation is small.

The transverse polarisation $P_x$ may be quite larger than the longitudinal one, and it slightly depends on the light quark couplings. For example, for the experimentally favoured region $\phi_\ell \sim \pi/4$, one has $P_x \lesssim 0.08$. At the Tevatron, this polarisation is as easy to measure as the longitudinal one, and the only limitation is the available statistics. At the LHC, one can use the motion of the $\ttb$ pair in the laboratory frame to select a preferred direction among the two protons~\cite{Baumgart:2013yra}, or study averaged azimuthal distributions that are symmetric under the exchange of the two proton momenta~\cite{Boudjema:2009fz,Godbole:2010kr,Baumgart:2011wk}. (Analogously, the normal polarisation can be probed by azimuthal angle distributions~\cite{Godbole:2006tq}.) The exploration of the sensitivity of these measurements is beyond the scope of this work.

The main results from our analysis of the relation between $\afb$, $\afbl$ and the top polarisation can be summarised as follows. For a given $\ttb$ asymmetry, say $\Delta \afb>0$ for definiteness, the lepton asymmetry can lie in a somewhat wide range --- provided that the quantum interference effects are properly taken into account --- and it can be larger or smaller than the SM value, depending mainly on the chirality of the octet coupling to the top quark. Focusing on a given lepton asymmetry, say $\Delta \afbl > 0$, the top longitudinal polarisation $P_z$ in the helicity basis can be positive, negative or nearly zero, depending on whether the octet couplings to $u$ and $d$ are predominantly right-handed, left-handed or axial, respectively. And, independently of these couplings, there is a sizeable transverse polarisation $P_x$ unless the top coupling is axial, and it should be experimentally searched for.

\section*{Acknowledgements}
This work has been supported by MICINN project FPA2010-17915; by
Junta de Andaluc\'{\i}a projects FQM 101, FQM 03048 and FQM 6552; and by FCT project EXPL/FIS-NUC/0460/2013.

\appendix
\section{Appendix}

\begin{figure}[htb]
\begin{center}
\begin{tabular}{cc}
\includegraphics[height=7cm,clip=]{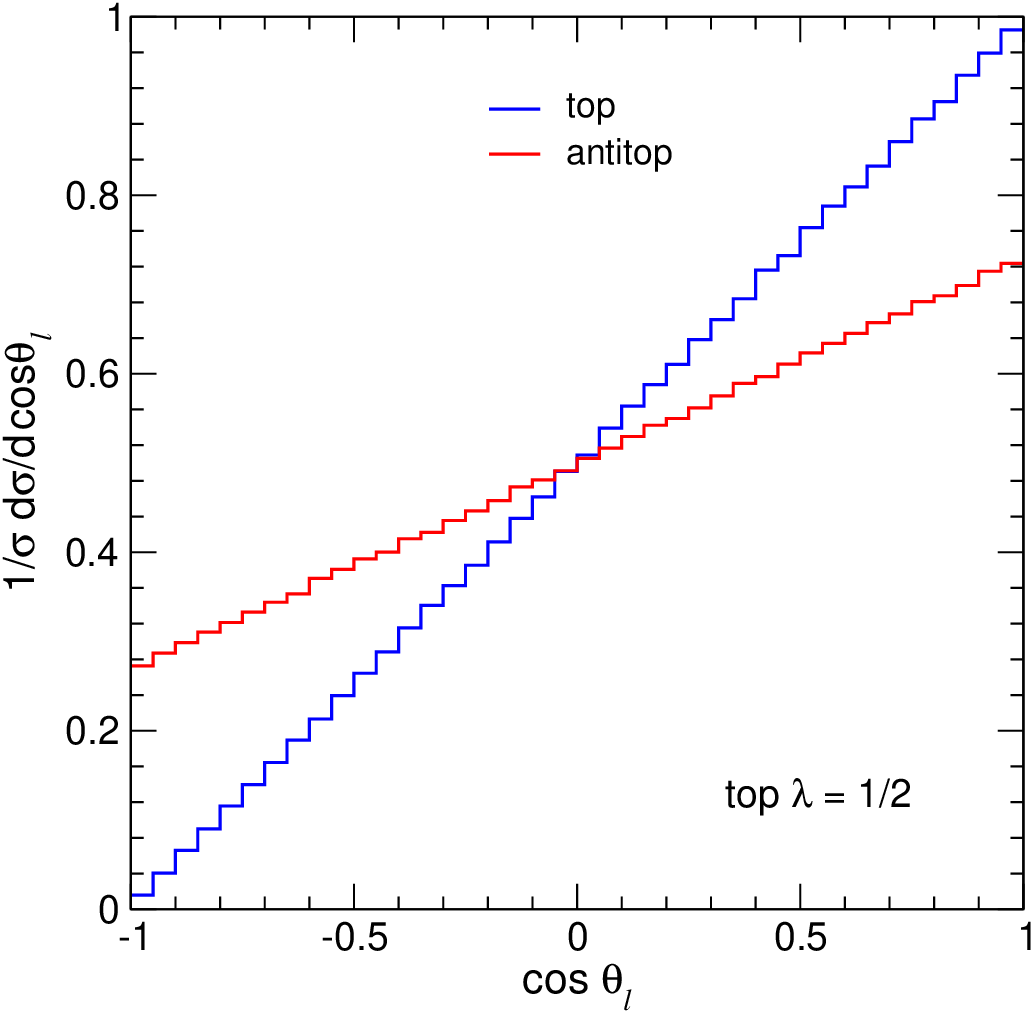} &
\includegraphics[height=7cm,clip=]{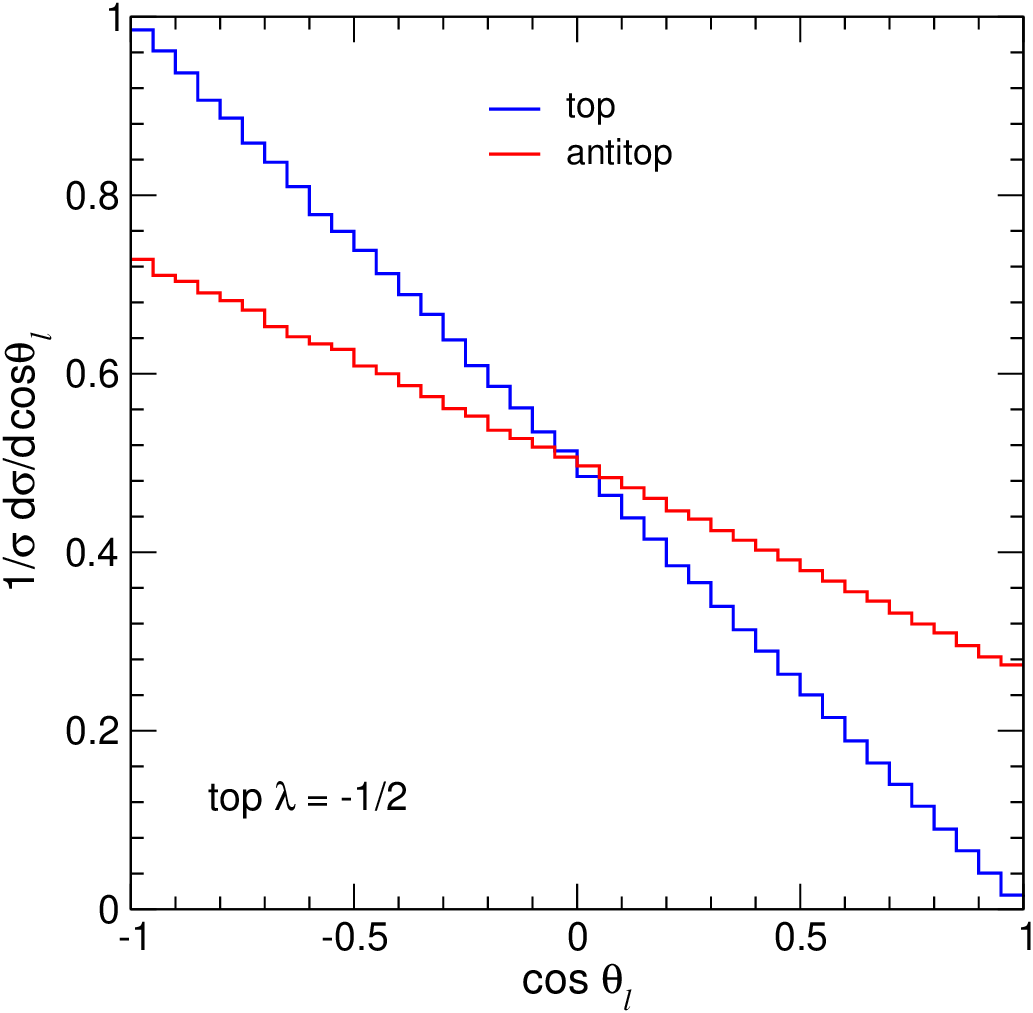} \\ 
\includegraphics[height=7cm,clip=]{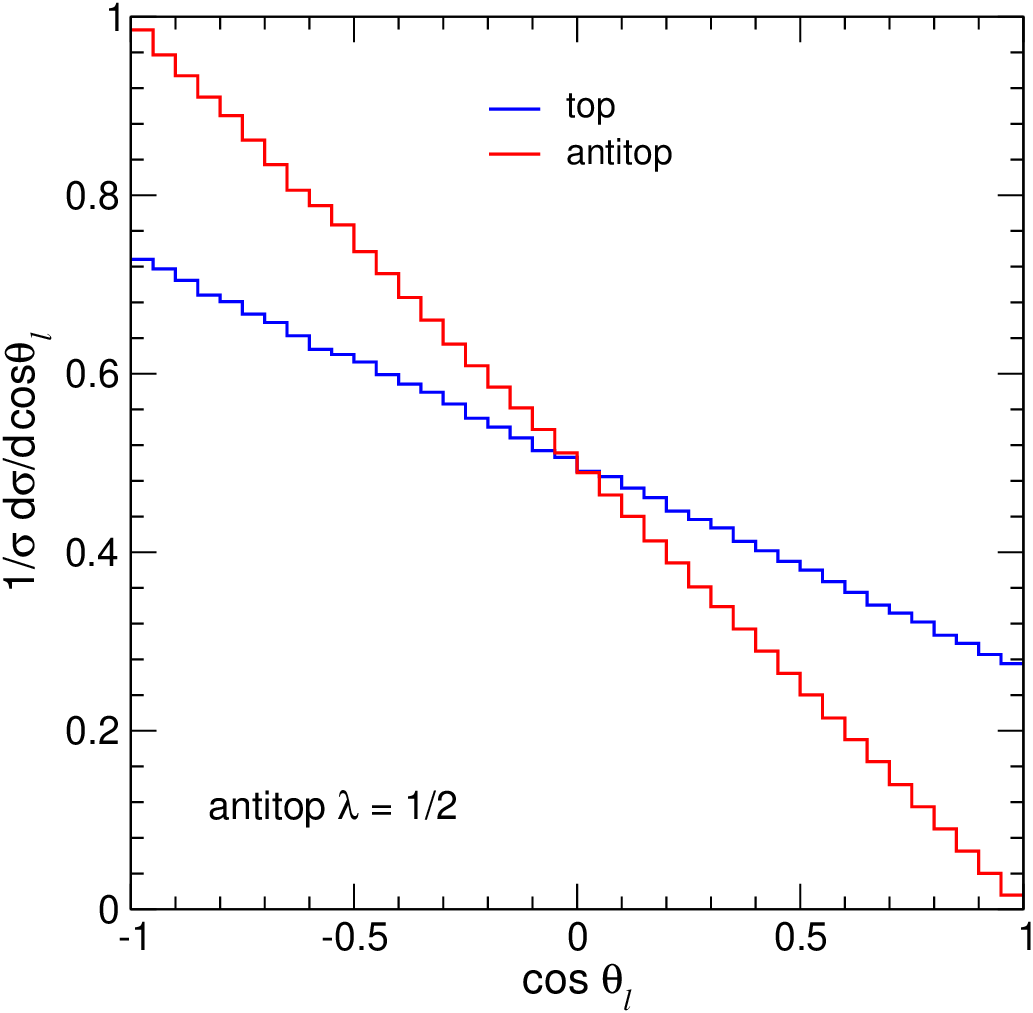} &
\includegraphics[height=7cm,clip=]{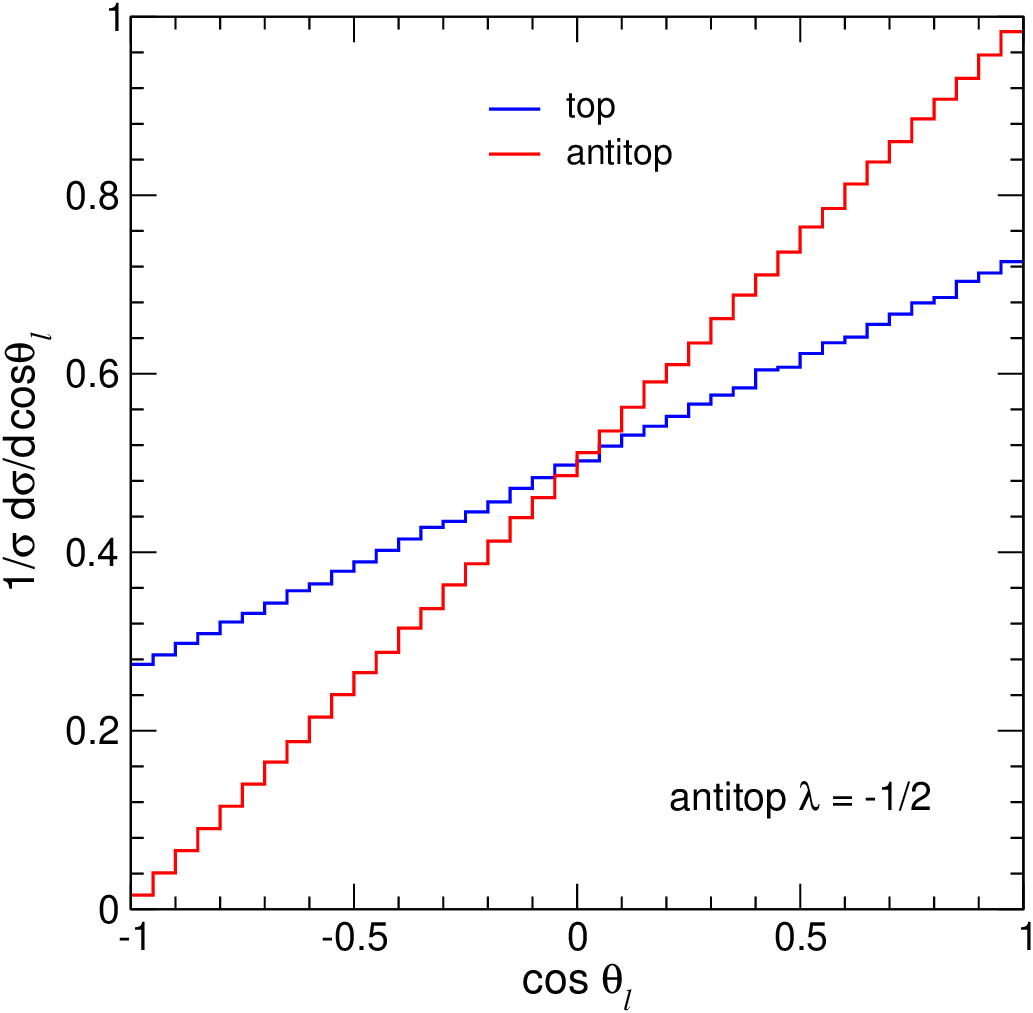} 
\end{tabular}
\end{center}
\caption{Angular distributions of the charged leptons in the (anti-)top rest frame, after selecting either top quarks (upper panels) or antiquarks (lower panels) of definite helicity.}
\label{fig:dist}
\end{figure}

The implementation of spin projectors in a Monte Carlo generator has its own interest for experimental analyses, and we discuss here its features in some detail. The angular distribution of the charged lepton in the top quark rest frame with respect to some $\hat z$ direction is given by
\begin{equation}
\frac{1}{\Gamma} \frac{d\Gamma}{d\cos\theta_\ell} = \frac{1}{2} ( 1 + P_z \alpha_\ell \cos \theta_\ell ) \,,
\label{ec:dist1}
\end{equation}
with $\alpha_\ell = 1$ for the positive charge leptons from the top decay $\ell^+ = e^+,\mu^+,\tau^+$ and $\alpha_\ell = -1$ for the negative charge ones from the antitop. Then, the distribution allows to measure the polarisation of the produced top (anti-)quarks. We test the helicity projectors in $\ppb \to \ttb$ within the SM by selecting different helicities $\lambda$ for the top and the antitop: (a) $\lambda = 1/2$ for $t$, no selection for $\bar t$; (b) $\lambda = -1/2$ for $t$, no selection for $\bar t$; (c) $\lambda = 1/2$ for $\bar t$, no selection for $t$; (b) $\lambda = -1/2$ for $\bar t$, no selection for $t$. The resulting lepton angular distributions are shown in Fig.~\ref{fig:dist}. 
When projecting a definite helicity for the top (antitop), the cross section halves and the distribution of the positive (negative) charged lepton is found as expected, with $P_z = \pm 1$. Moreover, selecting a helicity for one quark automatically polarises the companion quark, as it is expected from the spin correlation between them~\cite{Mahlon:1995zn}. The spin correlation in the helicity basis is
\begin{equation}
C = \frac{N(t_+ t_+) + N(t_- t_-) - N(t_+ t_-) - N(t_- t_+)}{N(t_+ t_+) + N(t_- t_-) + N(t_+ t_-) + N(t_- t_+)} \simeq - 0.45
\label{ec:C}
\end{equation}
and it is indeed observed that, when selecting $P_z = \pm 1$ for one of the quarks, the other quark acquires a polarisation $P_z = \mp 0.45$. Finally, we also test projecting definite helicities for both quarks, in which case the cross sections are
\begin{equation}
\sigma_{++} = 0.84~\text{pb} \,,\quad 
\sigma_{+-} = 2.22~\text{pb} \,,\quad
\sigma_{-+} = 2.22~\text{pb} \,,\quad
\sigma_{--} = 0.84~\text{pb} \,, 
\end{equation}
where the first and second subscript on $\sigma$ refer to the top and antitop helicity, respectively. These cross sections are in agreement with Eq.~(\ref{ec:C}).

\end{document}